\documentclass{aa} 

\usepackage{graphicx}
\usepackage{txfonts}
\usepackage[colorlinks=true,citecolor=blue,linkcolor=blue]{hyperref}

\begin{document}

   \title{The newborn black hole in GRB 191014C proves that it is \emph{alive}}

   \author{R.~Moradi
          \inst{1}\fnmsep\inst{2}\fnmsep\inst{3}
          \and
          J.~A.~Rueda\inst{1}\fnmsep\inst{2}\fnmsep\inst{4}\fnmsep\inst{5}\fnmsep\inst{6}
          \and
          R.~Ruffini\inst{1}\fnmsep\inst{2}\fnmsep\inst{7}
          \and
          Y.~Wang\inst{1}\fnmsep\inst{2}\fnmsep\inst{3}
          }
   \institute{ICRANet, Piazza della Repubblica 10, I--65122 Pescara, Italy\\
   \email{rahim.moradi@icranet.org, jorge.rueda@icra.it, ruffini@icra.it, yu.wang@icranet,org}
        \and
        ICRA, Dipartimento di Fisica, Sapienza Universit\`a di Roma, P.le Aldo Moro 5, I--00185 Rome, Italy
        \and
        INAF, Osservatorio Astronomico d'Abruzzo,Via M. Maggini snc, I-64100, Teramo, Italy
        \and
        ICRANet-Ferrara, Dipartimento di Fisica e Scienze della Terra, Universit\`a degli Studi di Ferrara, Via Saragat 1, I--44122 Ferrara, Italy
        \and
        Dipartimento di Fisica e Scienze della Terra, Universit\`a degli Studi di Ferrara, Via Saragat 1, I--44122 Ferrara, Italy
        \and
        INAF, Istituto de Astrofisica e Planetologia Spaziali, Via Fosso del Cavaliere 100, 00133 Rome, Italy
        \and
        INAF, Viale del Parco Mellini 84, 00136 Rome  Italy
        }

\date{Received Month day, year; accepted Month day, year}


\abstract{
{
A multi-decade theoretical effort has been devoted to finding an efficient mechanism to use the rotational and electrodynamical extractable energy of a Kerr-Newman black hole (BH), to power the most energetic astrophysical sources such as gamma-ray bursts (GRBs) and active galactic nuclei (AGN). We show an efficient general relativistic electrodynamical process which occurs in the ``{inner engine}'' of a binary driven hypernova (BdHN). The {inner engine} is composed of a rotating Kerr BH of mass $M$ and dimensionless spin parameter $\alpha$, a magnetic field of strength $B_0$ aligned and parallel to the rotation axis, and a very low-density ionized plasma. Here, we show that the gravitomagnetic interaction between the BH and the magnetic field induces an electric field that accelerates electrons and protons from the environment to ultrarelativistic energies emitting synchrotron radiation. We show that in GRB 190114C the BH of mass $M = 4.4~M_\odot$, $\alpha= 0.4$, and $B_0 \approx 4\times 10^{10}$~G can lead to a high-energy ($\gtrsim$GeV) luminosity of $10^{51}$~erg~s$^{-1}$. The \textit{inner engine} parameters are determined by requiring 1) that the BH extractable energy explains the GeV and ultrahigh-energy emission energetics, 2) that the emitted photons are not subjected to magnetic-pair production, and 3) that the synchrotron radiation timescale agrees with the observed high-energy timescale. We find for GRB 190114C a clear jetted emission of GeV energies with a semi-aperture angle of approximately $60^\circ$ with respect to the BH rotation axis.}
}


\titlerunning{The newborn black hole in GRB 190114C}
\authorrunning{Moradi, et al.}

\maketitle

\section{Introduction}\label{sec:1}

\begin{table*}[h]
{
    \centering
    \begin{tabular}{l|l|l|l}
        & cgs-Gaussian & geometric & cgs to geometric\\
        \hline
        $M$ & g & cm & $G/c^2 = 7.425\times 10^{-29}$~cm~g$^{-1}$\\
        $Q$ & cm$^{3/2}$~g$^{1/2}$~s$^{-1}$ (statC) & cm & $G^{1/2}/c^2 = 2.874\times 10^{-25}$~cm$^{-1/2}$~g$^{-1/2}$~s\\
        $J$ & g\,cm$^2$\,s$^{-1}$ & cm$^2$ & $G/c^3 = 2.477\times 10^{-39}$~g$^{-1}$~s\\
        $\Phi$ & cm$^2$ g s$^{-2}$ (erg) & cm & $G/c^4 = 8.261\times 10^{-50}$~cm$^{-1}$~g$^{-1}$~s$^2$ \\
        $\phi$ & cm$^{1/2}$~g$^{1/2}$~s$^{-1}$ (statV) & cm$^0$ & $G^{1/2}/c^2 = 2.874\times 10^{-25}$~cm$^{-1/2}$~g$^{-1/2}$~s\\
        $E$ & cm$^{-1/2}$~g$^{1/2}$~s$^{-1}$ (statV/cm) & cm$^{-1}$ & $G^{1/2}/c^2 = 2.874\times 10^{-25}$~cm$^{-1/2}$~g$^{-1/2}$~s\\
        $B$ & cm$^{-1/2}$~g$^{1/2}$~s$^{-1}$ (gauss, G) & cm$^{-1}$ & $G^{1/2}/c^2 = 2.874\times 10^{-25}$~cm$^{-1/2}$~g$^{-1/2}$~s\\
        \hline
    \end{tabular}
    \caption{{Units of the relevant  physical quantities used in this article in the cgs-Gaussian and geometric system of units.} Notation: $M$ mass, $Q$ charge, $J$ angular momentum, $\Phi$ electric potential energy, $\phi$ electric potential, $E$ electric field, $B$ magnetic field. We use length (cm) as the base unit in the geometric system.}
    \label{tab:units}
    }
\end{table*}

Rotating black holes (BHs) have  traditionally been described by the Kerr \citep{1963PhRvL..11..237K} and the Kerr-Newman metrics \citep{1965JMP.....6..918N} which assume three conditions:  (i) they are in a matter vacuum, (ii) they are embedded in an asymptotically flat spacetime, and (iii) they fulfill global stationarity. Under these conditions, BHs are just a sink of energy, {namely} ``{dead BHs}.''  The discovery of the reversible and irreversible transformations in both these spacetimes \citep{1970PhRvL..25.1596C, 1971PhRvD...4.3552C} opened the conceptual possibility of extracting both rotational and electromagnetic energy from a Kerr-Newman BH. These results also led  to the asymptotic mass-energy formula relating the mass $M$ of a BH to   three independent parameters, the irreducible mass $M_{\rm irr}$, the charge $Q$, and the angular momentum $J$, soon confirmed by \citet{1972CMaPh..25..152H}. The perspective that up to $50\%$ of the mass-energy of a Kerr-Newman BH could be extracted directed the attention to the alternative view of  ``\textit{alive BHs}'' whose extractable energy could be used as an astrophysical source (see, e.g., ``Introducing the black hole'' \citealp{Ruffini:1971bza} and ``On the energetics of black holes'' by R. Ruffini in ~\citealp{DeWitt:1973uma}).

Since then an efficient process  has been sought that is able to power the most energetic astrophysical sources, gamma-ray bursts (GRBs), and active galactic nuclei (AGNs), using the \textit{extractable} energy from a BH. The theoretical framework has been constantly evolving  \citep[see, e.g.,][for a review on this topic]{2019Univ....5..125T}. As we  show in this paper, the recent discovery of the birth of a BH {in GRB 130427A} \citep{2019ApJ...886...82R, 2020EPJC...80..300R} demonstrates that the Kerr BH harbored in the \textit{inner engine} of this source is {indeed} an enormous source of giga-electron volt (GeV) energy. The main topic of this article is to reach a deeper understanding of the process of rotational energy extraction by further identifying  the astrophysical setting,  the boundary conditions, and  the basic new physical laws that allow this process to become observable. Specifically, for the case of GRB 190114C, our goal is to infer  the values of the independent physical component,  the spectral distribution of the high-energy GeV emission, and the geometrical properties of the GeV and ultrahigh-energy emissions.

\textit{Astrophysical Setting}

Our approach is based on the binary-driven hypernova (BdHN) model of long GRBs \citep{2012ApJ...758L...7R, 2014ApJ...793L..36F, 2016ApJ...833..107B, 2019ApJ...886...82R}.
The BdHN progenitor is a binary system composed of a carbon-oxygen (CO) star and a neutron star (NS) companion. The collapse of the iron core of the evolved CO star forms a newborn NS ($\nu$NS) at its center and expels the stellar outermost layers, hence leading to a {supernova (SN)} explosion. The {SN} ejecta produces a hypercritical accretion process both onto the $\nu$NS and onto the NS companion. For very compact binaries (orbital period on the order of $5$~min), the NS companion  reaches the critical mass rapidly (a few seconds), undergoes gravitational collapse, and  forms a rotating BH. We have called these long GRBs in which there is BH formation, BdHNe of type I (BdHN I). Their isotropic energy release is in the range $10^{53}$--$10^{54}$~erg. Numerical simulations of the above process in one, two, and three dimensions have been presented in \citet{2014ApJ...793L..36F}, \citet{2015ApJ...812..100B}, and \citet{2016ApJ...833..107B, 2019ApJ...871...14B}, respectively. Only a fraction of BdHNe form BHs (380 BdHNe I have been identified; see \citealp{2021MNRAS.tmp..868R}). In progenitors with longer binary periods, on the order of hours, no BHs are formed; the outcome is a binary NS with long GRBs in the range $10^{51}$ to $10^{53}$~erg \citep{2019ApJ...874...39W}. For even longer binary periods, on the order of days,  even less energetic long GRBs are encountered, the BdHNe III, for example the case of GRB 060218 (Liang, et al., in preparation).

{\textit{Associated boundary conditions}}

We now return to GRB 190114C. It has already been shown that the collapse of the CO star, which triggers the complete GRB process in the presence of a binary NS companion, leads to a SN creating an additional NS (i.e., the $\nu$NS). The SN process is observed earlier at X-ray (up to a few keV) and soft gamma-ray (up to a few MeV) wavelengths,  and it has been referred to as ``\textit{SN-rise}'' \citep[see, e.g.,][]{2019ApJ...874...39W}. For GRB 190114C this occurs in the rest-frame interval $t_{\rm rf}\lesssim 1.99$~s. It carries an energy of  $E_{\rm SN-rise} = 2.82\times 10^{52}$~erg and is characterized by a blackbody plus cutoff power-law spectrum \citep{2019arXiv191012615L}. The short duration of the \textit{SN-rise} finds a natural explanation in the BdHN model. In a BdHN I the companion NS is separated at only $10^{10}$--$10^{11}$~cm (i.e., about $1$~light-second) from the CO star, implying that only the first spike becomes observable before the expanding SN ejecta triggers the hypercritical accretion process onto the NS companion \citep[see, e.g.,][]{2019ApJ...871...14B, 2019ApJ...874...39W}.

The newborn BH is embedded in the magnetic field inherited from the NS \citep{2020ApJ...893..148R}, and sits at the center of a {cavity} of very low density \citep[see][for numerical simulations]{2019ApJ...883..191R} of material from the SN ejecta. For GRB 190114C such a density has been  estimated to be on the order of $10^{-14}$~g~cm$^{-3}$. The cavity is carved during the accretion and subsequent gravitational collapse of the NS leading to the BH. The magnetic field remains anchored to material that did not participate in the BH formation (see \citealp{2020ApJ...893..148R} for a detailed discussion on the magnetic field around the newborn Kerr BH in a BdHN I).

The Kerr BH in the \textit{cavity} is therefore not isolated and acts in conjunction with a test magnetic field of strength $B_0$, aligned with the BH rotation axis. An additional important feature is that there is no vacuum surrounding the BH. As we show in this article, a fully ionized, very low-density plasma is essential to allow the electrodynamical performance of the energy extraction process by the \textit{inner engine}, which is necessarily non-stationary.

The operation procedure of the \textit{inner engine} leads the mass and spin of the BH to decrease as functions of time, while the BH irreducible mass ($M_{\rm irr}$) remains constant. The electrons accelerate to ultrahigh energies at the expense of the BH extractable energy\footnote{We use cgs-Gaussian units throughout, unless otherwise specified. Careted symbols stand for quantities in geometric units; for example, $\hat{M}\equiv G M/c^2$ denotes geometric mass. See Table~\ref{tab:units} for details on the units and conversion factors between the cgs-Gaussian and geometric systems of units.}
\begin{equation}\label{eq:Eextr}
{
    E_{\rm extr}\equiv (M-M_{\rm irr}) c^2,
    }
\end{equation}
obtainable from the BH mass-energy formula \citep{1970PhRvL..25.1596C, 1971PhRvD...4.3552C, 1971PhRvL..26.1344H}
\begin{equation}
\label{eq:Mbh}
M^2 = \frac{c^2}{G^2}\frac{J^2}{4 M^2_{\rm irr}}+M_{\rm irr}^2,
\end{equation}
where $J$ and  $M$ are respectively the angular momentum and the mass of the BH.
%

 {\textit{Physical laws of the GRB high-energy engine operation}}

As we explain in this article, using the mathematical role of the Papapetrou-Wald solution \citep{1966AIHPA...4...83P, 1974PhRvD..10.1680W},  a profound change of paradigm in relativistic astrophysics { has been made possible by the \textit{inner engine} \citep{2019ApJ...886...82R}, namely the introduction} of the effective charge given by the product of $J$ and $B_0$:
\begin{equation}\label{eq:Qeff}
    Q_{\rm eff} = \frac{G}{c^3} 2 J B_0.
\end{equation}
This effective charge originates from the gravitomagnetic interaction of the Kerr BH with the surrounding magnetic field, left over by the collapse of the accreting NS to the BH still rooted in the surrounding material \citep[see, e.g.,][]{2020ApJ...893..148R}. The existence of this effective charge finally explains the success of utilizing the concept of a Kerr-Newman BH as a temporary step to approach the analysis of quantum electrodynamical processes in the field of a rotating BH \citep[see, e.g.,][]{1975PhRvL..35..463D}.

{
We are now able {to elaborate, with the use of quantum electrodynamics and general relativity, a novel and physically more complete treatment of the GRB high-energy engine in a globally neutral system, therefore satisfying} Eq.~(\ref{eq:Mbh}). Starting from these general premises, the main focus of this article is the role of the newborn BH in giving origin to the GeV emission observed by Fermi-LAT {in the context of the \textit{inner engine} of a BdHN I. In Sect.~\ref{sec:2} we mathematically describe the electromagnetic field surrounding the Kerr BH following the Papapetrou-Wald solution of the Einstein-Maxwell equations \citep{1966AIHPA...4...83P,1974PhRvD..10.1680W}. Section~\ref{sec:3} summarizes the operation of the \textit{inner engine}, including its energy budget and electric potential energy available for the acceleration of charged particles around the BH. The particle motion along the BH rotation axis and its relation to the \textit{inner engine} contribution to ultrahigh-energy cosmic rays (UHECRs) is presented in Sect.~\ref{sec:4}. In Sect.~\ref{sec:5} we estimate the energy loss by synchrotron radiation for electrons moving outside the BH rotation axis. We obtain there the typical electron Lorentz factor, the corresponding pitch angles leading to high-energy ($\gtrsim$GeV) photons{, and the radiation timescale}. Section~\ref{sec:6} presents an estimate of the energy and angular momentum extracted to the Kerr BH in the emission process, in the radiation timescale, implied by the BH mass-energy formula. In Sect.~\ref{sec:7} we present our method of inferring the \textit{inner engine} parameters, namely the BH mass and spin, and the magnetic field strength {from} the three conditions required (the observed high-energy emission covered by the extractable energy of the BH, the observed high-energy luminosity  equal to the synchrotron radiation value, the emitted high-energy photons able to freely escape from the system). In Sect.~\ref{sec:8} we apply this framework to the case of GRB 190114C obtaining the corresponding \textit{inner engine} parameters. Section~\ref{sec:9} is dedicated to a comparison of our results with previous literature results. Finally, we outline the conclusions in Sect.~\ref{sec:10}.}

\section{{Electric and magnetic fields around the BH}}\label{sec:2}

\begin{figure*}
    \centering
    \includegraphics[width=0.6\hsize,clip]{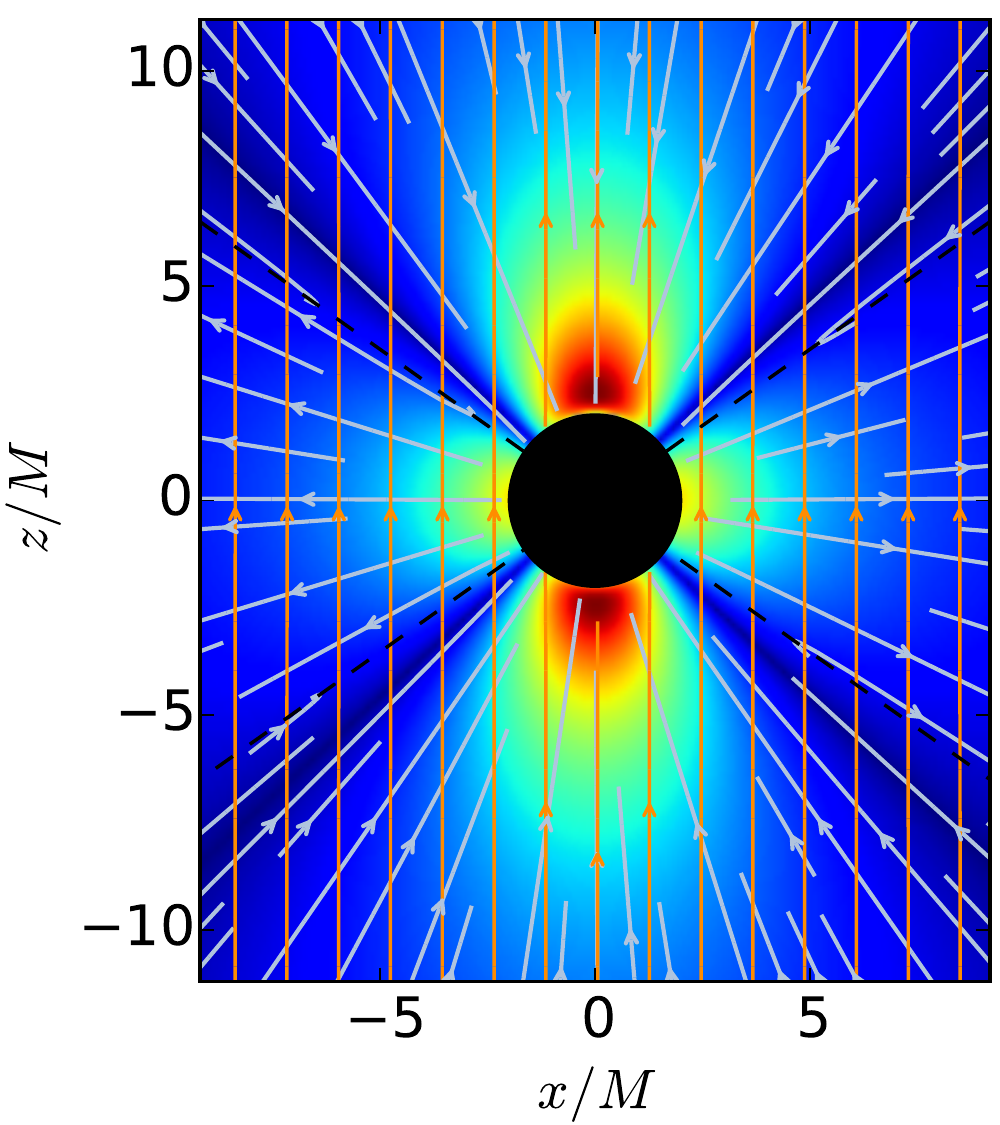}
    \caption{Electric (blue lines) and magnetic (golden lines) field lines of the Papapetrou-Wald solution in the xz-plane in Cartesian coordinates. The BH spin parameter is   set to $a/M = 0.3$ and the magnetic field and the BH spin are aligned and parallel. The background is a density-plot of the electric field energy density which is decreasing from red to blue. The BH horizon is the black disk. Distances are in units of $M$ and the fields in units of $B_0$. Outward electron acceleration occurs in the region limited by the dashed black lines, i.e., where the electric field is inwardly directed. In the northern hemisphere it covers spherical polar angles (measured clockwise) $-\theta_\pm<\theta<\theta_\pm$, where $\theta_{\pm} \approx 55^\circ$. By equatorial symmetry, in the southern hemisphere, it covers $\pi-\theta_\pm<\theta<\pi+\theta_\pm$.}
    \label{fig:fieldlines}
\end{figure*}

Following the considerations presented in \citet{2015ApJ...798...10R} and \citet{2019ApJ...886...82R} corresponding to GRB 130427A, we turn to a quantitative estimate of the \emph{inner engine} via a solution of the Einstein-Maxwell equations of a Kerr BH embedded in a test, asymptotically aligned, uniform magnetic field \citep{1966AIHPA...4...83P,1974PhRvD..10.1680W}, hereafter the  Papapetrou-Wald solution. 

The BH rotation and the aligned magnetic field induce an electric field {with the following radial and polar components:
\begin{subequations}\label{eq:Efieldzamo}
\begin{align}
E_{\hat{r}} &= \frac{B_0 \hat{a} \hat{M}}{\Sigma^2 A^{1/2}} \left[2 r^2 \sin^2\theta\,\Sigma - (r^2 + \hat{a}^2)(r^2-\hat{a}^2\cos^2\theta)(1+\cos^2\theta) \right],\\
E_{\hat{\theta}} &= B_0 \hat{a} \hat{M}\,\frac{\Delta^{1/2}}{\Sigma^2 A^{1/2}} 2\,r\,\hat{a}^2 \sin\theta \cos\theta\,(1+\cos^2\theta).
\end{align}
\end{subequations}
The magnetic field components are
\begin{subequations}\label{eq:Bfieldzamo}
\begin{align}
B_{\hat{r}} &= \frac{B_0 \cos\theta}{\Sigma^2 A^{1/2}} \left\{ (r^2+\hat{a}^2)\Sigma^2 -2 \hat{M} r \hat{a}^2 [2 r^2 \cos^2\theta + \hat{a}^2(1+\cos^4\theta)] \right\},\\
B_{\hat{\theta}}&= -\frac{\Delta^{1/2}}{\Sigma^2 A^{1/2}} B_0 \sin\theta\,[\hat{M} \hat{a}^2 (r^2 - \hat{a}^2\cos^2\theta)(1+\cos^2\theta) + r \Sigma^2],
\end{align}
\end{subequations}
where $\Sigma=r^2 + \hat{a}^2\cos^2\theta$, $\Delta=r^2-2\hat{M}r+\hat{a}^2$, $A = (r^2 + \hat{a}^2)^2-\Delta \hat{a}^2 \sin^2\theta$, and  {$\hat{M}=G\,M/c^2$ and $\hat{a}=\hat{J}/\hat{M} = (G J/c^3)/(G M/c^2) = J/(M c)$ are respectively  the geometric mass and specific angular momentum of the BH (see Table~\ref{tab:units})}.

{We here use the \textit{locally non-rotating} observer, also known as zero angular momentum observer (ZAMO; see  \citealp{1970ApJ...162...71B, 1972ApJ...178..347B}). Therefore, the electromagnetic field components (\ref{eq:Efieldzamo}--\ref{eq:Bfieldzamo}) differ from those presented in \citet{2019ApJ...886...82R}, where the Carter's observer was used.
}

{
For moderate dimensionless spin values, $\alpha \lesssim 0.7$, where $\alpha \equiv \hat{a}/\hat{M}= c J/(G M^2)$, the electric and magnetic fields are accurately represented by the {first-order expansion ($\alpha\ll 1$ or $\hat{a}\ll\hat{M}$)}:
\begin{subequations}\label{eq:EMslow}
\begin{align}
    E_{\hat{r}} &\approx -\frac{B_0 \hat{a} \hat{M}}{r^2}(3 \cos^2\theta-1),\quad E_{\hat{\theta}} \approx 0,\\
    B_{\hat{r}} &\approx B_0 \cos\theta,\quad     B_{\hat{\theta}} \approx -B_0 \sqrt{1-\frac{2 \hat{M}}{r}} \sin\theta.
\end{align}
\end{subequations}
}
Thus, the electric field is mainly radial and inwardly directed. {The electric field decreases with the square of the distance; therefore, it is maximum at the BH horizon, $r_+=\hat{M}(1+\sqrt{1-\alpha^2})$, and on the rotation axis $\theta = 0$, so $E_{\hat{r},\rm max} = -2 B_0 \hat{a}/r_+^2 = -\alpha B_0/2$. The electric field vanishes for $3 \cos\theta_\pm -1 = 0$. Therefore, it is inwardly directed in the northern hemisphere for spherical polar angles (measured clockwise) $-\theta_\pm<\theta<\theta_\pm$, where $\theta_{\pm} = \arccos(\sqrt{3}/3) \approx 55^\circ$ (see Eqs.~\ref{eq:EMslow} and Fig.~\ref{fig:fieldlines}). Because of the equatorial symmetry it also points inward in the southern hemisphere for $\pi-\theta_\pm<\theta<\pi+\theta_\pm$. In these regions, electrons are outwardly accelerated. In the remaining regions the electric field reverses sign, becoming outwardly directed (see Fig.~\ref{fig:fieldlines}). The value of $\theta_\pm$ is indeed accurately given by the slow-rotation approximation; for instance, a numerical calculation shows that $\theta_\pm\approx 54.74^\circ$--$59.76^\circ$ for $\alpha = 0.01$--$0.99$.} {We show below in this article that, the electrons located in these northern and southern hemisphere cones of semi-aperture angle of $\approx 60^\circ$, are outwardly accelerated with the appropriate pitch angles leading to GeV photons (see Sec.~\ref{sec:5} and Fig.~\ref{fig:chivsegamma} for details). Clearly, being anisotropic, this ``jetted'' emission is not always visible. This feature has been crucial for the inference of the morphology of the BdHN I from the high-energy (GeV) data of long GRBs \citep{2021MNRAS.tmp..868R}.}

It can be also seen that the magnetic field is everywhere nearly aligned with the BH rotation axis;  at any distance we have $B_z \gg B_x\sim B_y$, and at distances $r\gg 2 \hat{M}$ it is perfectly aligned (i.e., $B_x \to 0$, $B_y\to 0$, and $B_z\to B_0$). All these features can be seen in Fig.~\ref{fig:fieldlines}, which shows the electric and magnetic field lines given by the general expressions given by Eqs.~(\ref{eq:Efieldzamo})--(\ref{eq:Bfieldzamo}).

\section{Operation of the {inner engine}}\label{sec:3}

The operation of the \emph{inner engine} is based on three components naturally present in a BdHN I:
\begin{itemize}
    \item 
    the Kerr metric that describes the gravitational field produced by the newborn rotating BH;
    \item 
    an asymptotically uniform magnetic field around the newborn BH fulfilling the Papapetrou-Wald solution (see Sect.~\ref{sec:2});
    \item
    a very low-density plasma around the newborn BH composed of ions and electrons of $10^{-14}$~g~cm$^{-3}$ \citep{2019ApJ...883..191R}.
\end{itemize}

The \emph{inner engine} operation follows these precise steps:
\begin{enumerate}
    \item 
    The magnetic field and the BH rotation induce an electric field as given by the Papapetrou-Wald solution (see Sect.~\ref{sec:2}). For an aligned and parallel magnetic field to the BH spin, the electric field is nearly radial and inwardly directed at, and about, the BH rotation axis within an angle $\theta_\pm$ (see Fig.~\ref{fig:fieldlines}).
    \item
    The induced electric field accelerates electrons outwardly. The number of electrons that can be accelerated is set by the energy stored in the electric field \citet{2020EPJC...80..300R}:
\begin{subequations}\label{eq:Equantum}
\begin{align}
    {\cal E} &\approx \frac{1}{2}E_{\hat{r}}^2 r_+^3 = \hbar\,\Omega_{\rm eff},\\
    \Omega_{\rm eff} &= 4\left(\frac{m_{\rm Pl}}{m_n}\right)^8\left(\frac{B_0^2}{\rho_{\rm Pl}}\right)\alpha\,\Omega_+.
\end{align}
\end{subequations}
    Here $\Omega_+= c^2\partial M/\partial J = c\,\alpha/(2\,r_+)$ is the so-called BH angular velocity; $m_n$ the neutron mass; and $\rho_{\rm Pl}\equiv m_{\rm Pl} c^2/\lambda_{\rm Pl}^3$, $\lambda_{\rm Pl}=\hbar/(m_{\rm Pl} c),$ and $m_{\rm Pl}=\sqrt{\hbar c/G}$ are respectively the Planck energy-density, length, and mass. These expressions evidence the nature of the underlying physical process generating the electric field and the BH horizon: the electrodynamics of the Papapetrou-Wald solution \citep{2019ApJ...886...82R}, the origin of its magnetic field from the binary NS companion \citep{2020ApJ...893..148R}, and the smooth formation of the BH from the induced  gravitational collapse process \citep{2012ApJ...758L...7R}. Additional details on the above formulation are  presented in \citet{2020EPJC...80..300R}.
    \item
    The maximum possible electron acceleration and energy is set by the electric potential energy difference from the horizon to infinity can be written as \citep{2020EPJC...80..300R}
    \begin{subequations}\label{eq:deltaphi}
    \begin{align}
    \Delta \Phi &= \frac{1}{c}e\,a\,B_0, = \hbar\,\omega_{\rm eff},\\
    \omega_{\rm eff} &= \frac{G}{c^4}4\left(\frac{m_{\rm Pl}}{m_n}\right)^4\,e\,B_0\,\Omega_+,
\end{align}
    \end{subequations}
    {where $a=J/M$}.
    \item 
    Along the polar axis radiation losses are absent (see below in Sect.~\ref{sec:4}), while at off-axis latitudes (see below in Sect.~\ref{sec:5}) the accelerated electrons emit synchrotron radiation. The radiation timescale $\tau_{\rm rad}$ must fulfill
    \begin{equation}
        \tau_{\rm rad} = \frac{{\cal E}}{L_{\rm GeV}},
    \end{equation}
    where $L_{\rm GeV}$ is the observed GeV luminosity.
    \item 
    After this, the energy ${\cal E}$ has been used and emitted. The process restarts with a new angular momentum $J = J_0- \Delta J$, being $\Delta J$ the angular momentum extracted to the Kerr BH by the event (see below Eq.~\ref{eq:dJ} in Sect.~\ref{sec:6}). 
    \item
    The above steps are repeated, with the same efficiency, if the density of plasma is sufficient, namely if the number of the particles is enough to cover the new value of the energy ${\cal E}$. Therefore, the \emph{inner engine} evolves in a sequence of \emph{elementary processes}, each emitting a well-defined, precise amount of energy.
\end{enumerate}

For the sake of example, let us chose fiducial parameters  $B_0=10^{11}$~G, $M=3 M_\odot$, and $\alpha=0.5$. In this case the {available} energy is ${\cal E}\approx 3.39\times 10^{37}$~erg, and the maximum energy that an accelerated electron can gain is {$\Delta \Phi \approx 1.06\times 10^7$~erg $=6.64\times 10^{18}$~eV}.

\section{Acceleration on the polar axis: Ultrahigh-energy cosmic rays}\label{sec:4}

Along the polar axis, $\theta=0$, the electric and magnetic fields only have the $z$-component and are thus parallel; see Eqs.~(\ref{eq:Efieldzamo})--(\ref{eq:Bfieldzamo}), or Eqs.~(\ref{eq:EMslow})). Since the electron is accelerated by the electric field, this implies that the electron pitch angle, which is  the angle between the electron's injection velocity (into the magnetic field) and the magnetic field, is zero. Consequently, no radiation losses (by synchrotron emission) occur for motion along the BH rotation axis.

The electrons accelerate outward gaining the total electric potential energy, $\Delta \Phi\sim 10^{18}$~eV. Therefore, the maximum number of electrons that the \emph{inner engine} can accelerate along the axis is 
\begin{equation}
    N_{\rm pole}=\frac{{\cal E}}{\Delta \Phi} = \frac{\Omega_{\rm eff}}{\omega_{\rm eff}}\sim 10^{31}.
\end{equation}
These ultrarelativistic electrons contribute to leptonic UHECRs. The timescale of this acceleration process along the polar axis is 
\begin{equation}\label{eq:taupole}
    \tau_{\rm pole} \equiv \frac{\Delta \Phi}{e E_{\hat{r}}\,c}\approx \frac{r_+}{c} = \frac{\alpha}{2 \Omega_+}\approx 10^{-5}\,\,{\rm s}.
\end{equation}
This implies that the \emph{inner engine} can accelerate electrons along the BH rotation axis at a rate
\begin{equation}\label{eq:dotNpole}
    \dot{N}_{\rm pole}\equiv \frac{N_{\rm pole}}{\tau_{\rm pole}}\sim 10^{36}\,\,{\rm s}^{-1},
\end{equation}
leading to a power 
\begin{equation}\label{eq:dotEpole}
    \dot{{\cal E}}_{\rm pole} = \dot{N}_{\rm pole}\Delta \Phi = \frac{{\cal E}}{\tau_{\rm pole}}\sim 10^{54}\,\,{\rm eV~s}^{-1}\approx 10^{42}\,\,{\rm erg~s}^{-1}.
\end{equation}
Since the electric and magnetic fields along the rotation axis (and nearly close to it) are parallel ({see Fig.~\ref{fig:fieldlines}}), the particles in that region are all accelerated (nearly) parallel to the BH rotation axis. Therefore, we do not expect the accelerated particles to have appreciable collisions able to reduce the above estimate of their maximum kinetic energy gain. {Therefore, $\dot{{\cal E}}_{\rm pole}$ given by Eq.~(\ref{eq:dotEpole}) is the maximum power available for UHECRs.} 

The extension of the considerations  presented here to very massive BHs and AGN, the role of the accretion disk in these galactic configurations, and the possibility of accelerating protons to produce UHECRs by the BH have started to be addressed \citep{2020EPJC...80..300R}. We compare and contrast in Table~\ref{tab:parameters} some of the  \emph{inner engine} physical properties applied to the case of GRB 190114C and to M87*.
\begin{table*}
    \centering
    {
    \begin{tabular}{c|c|c}
    \hline & GRB 190114C & AGN (M87*-like)\\
    \hline
    $M$ ($M_\odot$) & $4.4$ & $6.0\times 10^9$\\
    $\alpha$ & $0.4$ & $0.1$ \\
    $B_0$ (G) & $4.0\times 10^{10}$ & $10$\\
       $\tau_{\rm pole}$ & $4.33\times 10^{-5}$~s & $0.68$~d\\
       $\Delta \Phi$ (eV) & $3.12\times 10^{18}$ & $2.66\times 10^{17}$\\
       ${\cal E}$~(erg) & $7.02\times 10^{37}$ & $6.96\times 10^{44}$\\
       $\dot{{\cal E}}_{\rm pole}$~(erg~s$^{-1}$) & $1.62\times 10^{42}$ & $1.18\times 10^{40}$\\
       $\chi$ ($^\circ$)  & $0.1805$--$18.05$ & $0.0451$--$4.51$ \\ 
        $t_c$~(s)  & $1.45\times 10^{-16}$--$1.45\times 10^{-14}$ & $0.2939$--$29.39$ \\
        $L_{\rm GeV}$ (erg~s$^{-1}$) & $4.83\times 10^{51}$--$4.83\times 10^{53}$ & $2.37\times 10^{43}$--$2.37\times 10^{45}$\\
       \hline
    \end{tabular}
    }
    \caption{Some astrophysical properties of the \emph{inner engine} for GRB 190114C and AGN, in the latter adopting as a proxy M87* \citep{2020EPJC...80..300R}. The timescale of particle acceleration along the BH rotation axis $\tau_{\rm pole}$ is given by Eq.~(\ref{eq:taupole}); the maximum energy gained in such acceleration $\Delta \Phi$ is given by Eq.~(\ref{eq:deltaphi}). {The energy ${\cal E}$ available for acceleration and radiation} is given by Eq.~(\ref{eq:Equantum}). The {maximum power available for} acceleration {(i.e., to power UHECRs)} is $\dot{{\cal E}}_{\rm pole}$ and is given by Eq.~(\ref{eq:dotEpole}). {The pitch angle $\chi$ is computed from Eq.~(\ref{gas}) adopting the photon energy range $0.1$--$1$~GeV photons; the corresponding synchrotron radiation timescale $t_c$ is given by Eq.~(\ref{tcr}), and an estimate of the associated GeV luminosity, $L_{\rm GeV}\sim {\cal E}/t_c$, is also shown.} In both cases the corresponding \emph{inner engine} parameters (BH mass $M$, spin $\alpha$, and surrounding magnetic field strength $B_0$) have been fixed to explain the observed high-energy ($\gtrsim$~GeV) luminosity (see Sect.~\ref{sec:7} for the case of GRB 190114C and \citealp{2020EPJC...80..300R} for M87*).}
    \label{tab:parameters}
\end{table*}

\section{Acceleration at off-axis latitudes: Synchrotron radiation}\label{sec:5}

\begin{figure}
    \centering
    \includegraphics[width=\hsize,clip]{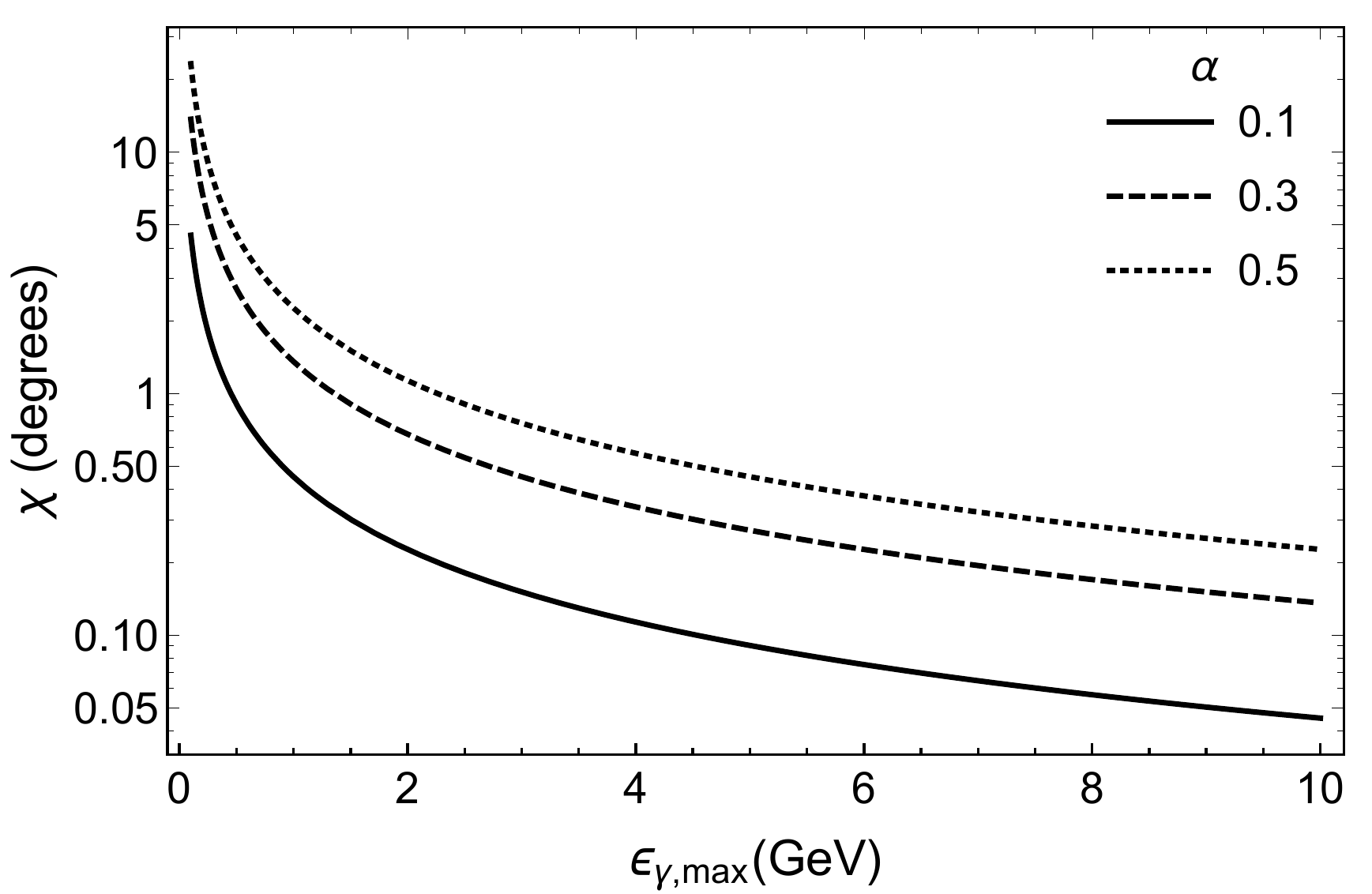}
    \caption{Pitch angle $\chi$ (in units of degrees) as a function of the photon critical energy $\epsilon_\gamma$ (in units of GeV) obtained from Eq.~(\ref{maxgev}). The focus of the plot for photon critical energy is in the range $0.1$--$10$~GeV. The solid, dashed, and dotted curves correspond to selected values of the BH spin parameter $\alpha = 0.1$, $0.3$, and $0.5$, respectively.}
    \label{fig:chivsegamma}
\end{figure}

For the present electric field, and assuming   radial motion, the dynamics of the electrons in the electromagnetic field, for $\gamma\gg 1$, is determined from \citep{1996ApJ...457..253D,2019ApJ...886...82R}
\begin{equation}
m_e c^{2}\frac{d\gamma}{dt}=e\frac{1}{2}\alpha B_{0}\,c-\frac{2}{3}e^4\frac{B_0^2\sin^2\left\langle \chi\right\rangle}{m_e^2 c^3}\gamma^2,\label{eqm}%
\end{equation}
where $e$ is the elementary charge, $\gamma$ is the electron Lorentz factor, $\left\langle \chi\right\rangle$ is the injection angle between the direction of electron motion and the magnetic field (the pitch angle), and $m_e$ is the electron mass. This equation is integrated assuming the electrons are injected near the horizon {(where the electric field strength is $\alpha B_0/2$)}, for selected values of the injection angle $\left\langle \chi\right\rangle$, with an initial Lorentz factor $\gamma=1$ at $t=0$. 

The synchrotron spectrum peaks roughly at the photon critical energy \citep[see, e.g.,][]{1975ctf..book.....L}
\begin{equation}\label{eq:egamma}
\epsilon_{\gamma}=\frac{3e\hbar}{2m_e\,c}B_{0}\sin\left\langle
\chi\right\rangle \gamma^2 = \frac{3}{2} m_e c^2 \beta\sin\left\langle
\chi\right\rangle \gamma^2,
\end{equation}
where in the last expression we   introduced $\beta=B_0/B_c$, with $B_c=m_e^2 c^2/(e\hbar)\approx 4.41\times 10^{13}$~G. Therefore, the synchrotron peak energy shifts from lower to higher energies (soft-to-hard spectral evolution) as the electron accelerates. {For example, the photon critical energy $\epsilon_{\gamma}$, for $\gamma \gtrsim 10^3$, a magnetic field $B_0 = 10^{11}$~G (so $\beta = 0.0023$), and a pitch angle $\chi = 10^\circ$ falls in the GeV regime}.

During the acceleration, the Lorentz factor increases linearly with time up to an asymptotic maximum value \citep[see][for details]{2019ApJ...886...82R}. This maximum value, set by the balance between the energy gain by acceleration in the electric field and energy loss by synchrotron radiation, is \citep{2019ApJ...886...82R}
\begin{equation}
\gamma_{\mathrm{max}}=\frac{1}{2}\left[\frac{3}{e^2/(\hbar
c)}\frac{\alpha}{\beta\sin^{2}\left\langle \chi\right\rangle
}\right]^{1/2},
\label{gas}%
\end{equation}
which defines the maximum electron energy $\epsilon_e=\gamma_{\rm max} m_e c^2$. Associated with $\gamma_{\rm max}$, by replacing Eq.~(\ref{gas}) into (\ref{eq:egamma}) we obtain the maximum peak energy of the spectrum \citep{2019ApJ...886...82R}
\begin{equation}\label{maxgev}
\epsilon_{\gamma,\mathrm{max}}=\frac{9}{8}\frac
{m_{e}c^{2}}{e^{2}/\hbar c}\frac{\alpha}{\sin\left\langle \chi
\right\rangle}\approx \frac{78.76}{\sin\left\langle \chi\right\rangle
}\alpha\,\, \mathrm{MeV},
\end{equation}
and the synchrotron cooling timescale $t=t_c$ for the above maximum photon critical energy is given by \citep{2019ApJ...886...82R}
\begin{equation}
t_c=\frac{\hbar}{m_e c^2}\frac{3}{\sin\left\langle \chi\right\rangle
}\left(\frac{e^2}{\hbar c}\alpha\,\beta^3\right)^{-1/2}.%
\label{tcr}
\end{equation}
For model parameters $\alpha = 0.5$ and $B_0=10^{11}$~G, photons of energy {$0.1$--$10$~GeV (typical photon energy range detected by the \textit{Fermi}-LAT) are emitted by electrons with pitch angles $\chi\approx 0.23$--$23^\circ$, and electron energy $\epsilon_e = 1.98 \times 10^{8}$--$1.98\times 10^{10}$~eV, radiating on a timescale of $t_c= 2.63\times 10^{-17}$--$2.63\times 10^{-15}$~s}. We show in Fig.~\ref{fig:chivsegamma} the pitch angle $\chi$ as a function of the maximum photon critical energy (spectrum peak energy) $\epsilon_{\gamma, \rm max}$, obtained from Eq.~(\ref{maxgev}), in the energy range $0.1$--$10$~GeV, and for three selected values of $\alpha$. {Figure~\ref{fig:pitchangles} shows the contours of constant $\chi$ for electrons moving in the electromagnetic field of the Papapetrou-Wald solution shown in Fig.~\ref{fig:fieldlines}.} {In particular, we show pitch angles for which electrons emit photons of GeV energies (see also Fig.~\ref{fig:chivsegamma}). It can be seen that this high-energy \textit{jetted} emission occurs within an effective opening angle $\theta_\pm \approx 60^\circ$. This anisotropic emission is essential to infer the BdHN I morphology from the GeV emission data of long GRBs \citep{2021MNRAS.tmp..868R}.}

\begin{figure*}[h!]
    \centering
    \includegraphics[width=0.75\hsize,clip]{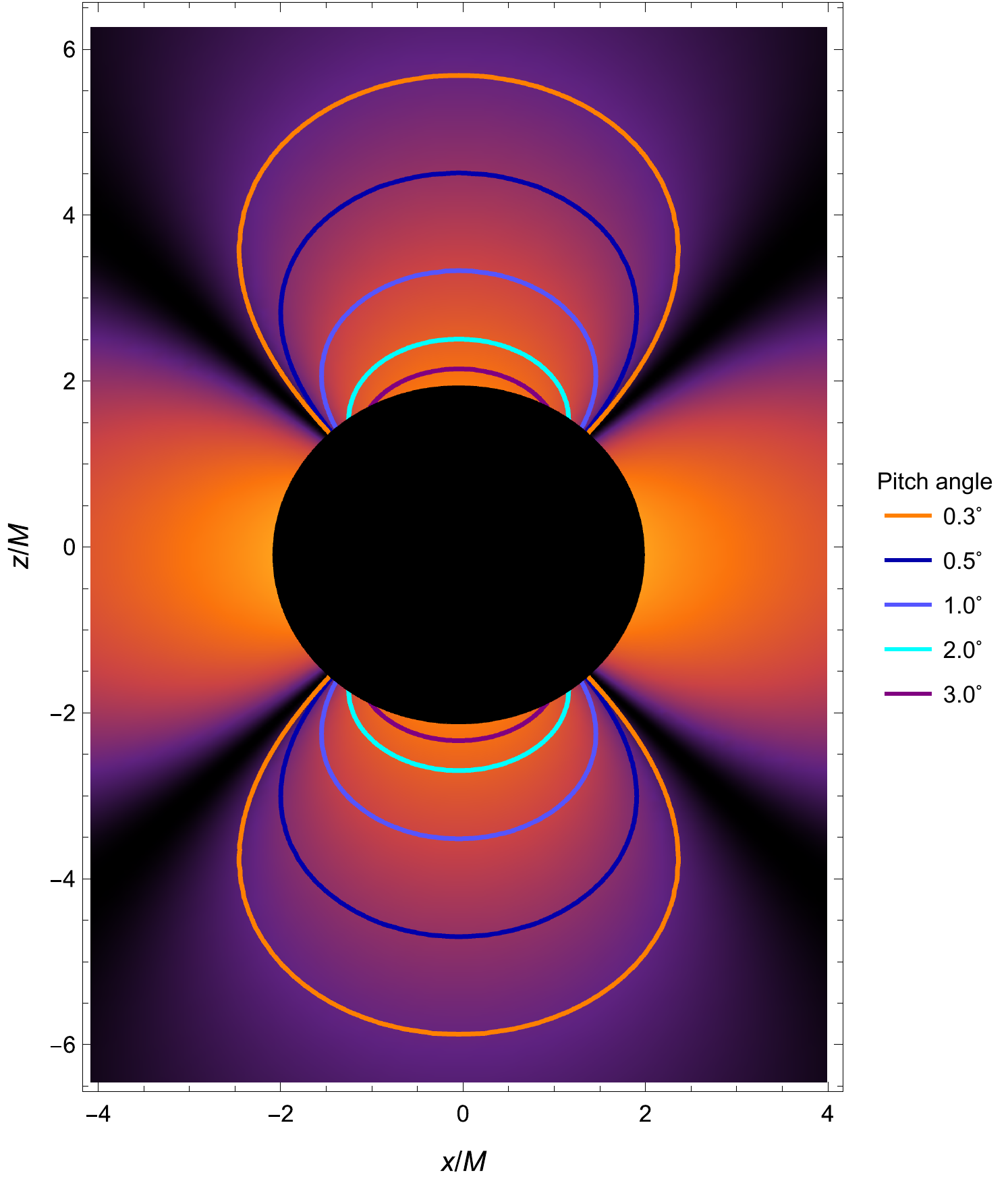}
    \caption{Contours of constant pitch angle ($\chi$) of electrons moving in the electromagnetic field of the Papapetrou-Wald solution shown in Fig.~\ref{fig:fieldlines}. For the present magnetic dominated case (|\vec{E}|/|\vec{B}| < 1), charged particles follow the magnetic field lines; therefore, $\sin\chi = |\vec{B}|^{-1}\sqrt{(\vec{E} + \vec{v}\times \vec{B}/c)^2 - (\vec{v}\cdot \vec{E})^2/c^2} \approx  \sqrt{|\vec{E}|^2/|\vec{B}|^2 - |\vec{E}_{||}|^2/|\vec{B}|^2}\approx |\vec{E}|/|\vec{B}|\sin\Psi$ (see, e.g., \citealp{2015AJ....149...33K}), where $\vec{E}_{||}$ is the electric field component parallel to the magnetic field, and $\Psi$ is the angle between $\vec{B}$ and $-\vec{E}$ (the minus sign is used because we are interested in the pitch angle of electrons). In the slow-rotation regime (see Eqs.~\ref{eq:EMslow}), $\sin\Psi \approx \sin\theta/(1-2 M \sin^2\theta/r)$, so $\sin\chi \approx |\vec{E}|/|\vec{B}|\sin\theta/(1-2 M \sin^2\theta/r)$. The BH is indicated by the   black disk. The background color map indicates the electric field energy density (lighter colors  means more intense). }
    \label{fig:pitchangles}
\end{figure*}

\section{{Energy and angular momentum extraction, {golden rule}, and} duration of the {inner engine} activity}\label{sec:6}

At the end of every elementary process, all the energy ${\cal E}$ has been emitted. The \emph{inner engine} restarts its operation with the same magnetic field of $B_0$, but with a new slightly smaller angular momentum $J = J_0-\Delta J$, being $\Delta J$ the angular momentum extracted by the {process}. From the BH mass-formula (\ref{eq:Mbh}), keeping the irreducible mass constant (i.e., $\Delta M_{\rm irr}=0$, and $\Delta M c^2 = {\cal E}$), we obtain a change in the BH angular momentum $\Delta J$ in each event:
\begin{subequations}\label{eq:dJ}
\begin{align}
    J_{\rm eff}&=  I_{\rm eff}\Omega_{\rm eff}, \quad I_{\rm eff} = M \left(\frac{2 G M_{\rm irr}}{c^2}\right)^2, \\
    \Delta J &= \frac{J_{\rm eff}}{J}\hbar,\\
    \Delta J &= \frac{1}{\bar{\Omega}}\Delta E,\quad \bar{\Omega}\equiv \frac{J}{I_{\rm eff}}, \quad \Delta E \equiv \mathcal{E}.
\end{align}
\end{subequations}
Here the last equation, a truly \textit{golden formula}, relates the energy radiated ({$\Delta E = {\cal E}$}) to the angular momentum extracted to the rotating BH ({$\Delta J$}).

For the fiducial parameters that we  used above, $M = 3~M_\odot$, $\alpha = 0.5$, and $B_0 = 10^{11}$~G, we have $J\approx 3.96\times 10^{49}$~g~cm$^2$~s$^{-1}$, $M_{\rm irr}\approx 2.9~M_\odot$, and $\Delta J\approx 1.0\times 10^{33}$~g~cm$^2$~s$^{-1}$, so a fractional change $\Delta J/J\approx 10^{-16}$, implying that the activity can last for thousands of years or more, providing there is ionized plasma to feed the \emph{inner engine}.

\section{Inference of the BH mass, spin, and surrounding magnetic field}\label{sec:7}

We require three physical and astrophysical conditions to obtain the three \emph{inner engine} parameters, the BH mass and spin, $M$ and $\alpha$, and the strength of the magnetic field surrounding the BH, $B_0$.
Following \citet{2019ApJ...886...82R}, who show that the use of solely the GeV emission data, after the ultrarelativistic prompt emission (UPE) phase \citep[see][]{2019arXiv191012615L}, is enough to determine the \emph{inner engine} parameters {(see Fig.~\ref{fig:GeVlum})}. In particular, we  show that this procedure serves to obtain a lower limit to the mass and spin of the BH. The most important point is that we obtain the value of the irreducible mass of the BH that is kept constant through the energy extraction process. This allows us to determine the time evolution of the BH mass and spin. This can be achieved by fulfilling the three following conditions.

\subsection{Condition 1}

First, we require that the rotational energy of the BH provides the energy budget for the observed GeV emission energetics, {
\begin{equation}\label{eq:Eext}
    E_{\rm extr}\geq E_{\rm GeV},
\end{equation}
which via Eqs.~(\ref{eq:Eextr}) and (\ref{eq:Mbh}) leads to the following inequality between $M$, $\alpha$ and $E_{\rm GeV}$}: 
\begin{equation}\label{eq:mu}
    M \geq \frac{1}{\eta}\frac{E_{\rm GeV}}{c^2},\quad \eta \equiv1-\sqrt{\frac{1 + \sqrt{1 - \alpha^2}}{2}}.
\end{equation}

{We recall that the maximum value of the efficiency parameter is $\eta_{\rm max} \approx 0.293$, which is attained for a maximally rotating BH, $\alpha_{\rm max}=1$. It is also important to recall that, by keeping the BH irreducible mass constant in the energy extraction process, we are inferring a lower limit to the BH mass. An increasing $M_{\rm irr}$ with time implies a higher BH mass to explain the same GeV energetics.}

\subsection{Condition 2}

We require that the GeV photons must be transparent to the magnetic $e^+e^-$ pair creation process. The attenuation coefficient for this process is (see \citealp{1983ApJ...273..761D} and Sect.~5 in \citealp{2019ApJ...886...82R})
\begin{equation}\label{eq:MFP}
    \bar{R}\sim0.23\frac{e^{2}}{\hbar c}\left( \frac{\hbar}{m_{e}c}\right)
^{-1}\beta\sin\left\langle \chi\right\rangle\exp\left(
-\frac{4/3}{\psi}\right),
\end{equation}
where $\psi = \beta\sin\left\langle
\chi\right\rangle\epsilon_\gamma/(2m_e c^2)$. Substituting Eq.~(\ref{maxgev}) into Eq.~(\ref{eq:MFP}), $\bar{R}$ becomes a function of $\epsilon_\gamma$ and the product $\alpha \beta$. For a given $\epsilon_\gamma$ and $\alpha$, the lower the magnetic field, the larger the mean free path $\bar{R}^{-1}$, as expected. When $\chi\ll 1$, the exponential term dominates, hence $\bar{R}^{-1}$ exponentially increases tending to become infinite. An order-of-magnitude estimate of the magnetic field can be obtained by requiring $\psi\ll 1$,
\begin{equation}\label{eq:Bmax1}
    \beta \ll \frac{16}{9}\frac{e^2}{\hbar c}\frac{1}{\alpha} \approx \frac{1.298\times 10^{-2}}{\alpha},\,\, {\rm or}\,\, B_0\ll \frac{5.728\times 10^{11}}{\alpha}\,\,{\rm G},
\end{equation}
which is independent of the photon peak energy. It should be noted that  this constraint already restricts the magnetic field to be undercritical ($\beta <1$), and as we shall see it is sufficient for explaining the GeV emission after the UPE phase. The constraint (\ref{eq:Bmax1}) is analogous to imposing a lower limit on $\bar{R}^{-1}$. For instance, adopting a photon energy of $0.1$ GeV, it can be checked that for $\alpha\,\beta = 1.298 \times 10^{-2}$, the mean free path is $\bar{R}^{-1}= 1.17\times 10^5$~cm. Lower values of $\alpha\,\beta$ lead to much larger values of $\bar{R}^{-1}$. It is very interesting that this value is comparable to $G M_\odot/c^2 \approx 1.477\times 10^5$~cm. Therefore, requesting a value of $\alpha\,\beta$ lower than the above-mentioned one, implies   having a mean free path that is  much larger than the BH horizon. Specifically, the high-energy photons are produced in the vicinity of the BH, but they can freely escape from the system. If we adopt as a fiducial value that $0.1$~GeV photons have a sufficiently large mean free path (e.g., $\bar{R}^{-1} \geq 10^{16}$~cm), we obtain \citep{2019ApJ...886...82R}
\begin{equation}\label{eq:Btransparency}
  \beta \leq \frac{3.737 \times 10^{-4}}{\alpha},\quad {\rm or}\quad B_0 \leq \frac{1.649\times 10^{10}}{\alpha}\,\,{\rm G}.
\end{equation}
That we are in the exponentially increasing part of the mean free path is evident by the fact that, by requesting a mean free path which is 11 orders of magnitude larger than the one implied by (\ref{eq:Bmax1}), our upper limit to the magnetic field is decreased less than one order of magnitude. {Therefore, our estimate of the magnetic field is not sensitive to the choice of the value of $\bar{R}^{-1}$, providing it satisfies $\gtrsim 10^{5}$~cm.} This implies that the magnetic field strength of the \emph{inner engine} is constrained to have a value, roughly speaking, in the range $10^{10}$--$10^{11}$~G {(see Fig.~\ref{fig:Bvsalpha})}.

\subsection{Condition 3}

The third condition (i.e., the closure equation)  is obtained by requesting that the timescale of the synchrotron radiation, the cooling time $t_c$ given by Eq.~(\ref{tcr}), be equal to the observed GeV emission timescale \citep{2019ApJ...886...82R}
\begin{equation}
    \tau_{\rm rad, 1} = \frac{{\cal E}_1}{L_{\rm GeV,1}},
\end{equation}
where ${\cal E}$ is the electrostatic energy available for the process (see  Eq.~(\ref{eq:Equantum})). The subscript ``1'' refers to quantities evaluated at the beginning of the transparency of the GeV emission (i.e., at the end of the UPE phase) at $t = t_{\rm rf, UPE}$ (see   Sect.~\ref{sec:8}). We refer to this as the first elementary impulsive event. Therefore, the third equation of the system is
\begin{equation}\label{eq:cond3}
  t_c \left(\left\langle\chi\right\rangle,\alpha,\beta\right)= \tau_{\rm rad, 1}\left(\mu,\alpha,\beta,L_{\rm GeV,1}\right),
\end{equation}
where $\mu=M/M_{\odot}$ and $M_{\odot}$ is the solar mass.

Therefore, having imposed these three conditions, we obtain the three \emph{inner engine} parameters from the system of equations (\ref{eq:mu}), (\ref{eq:Btransparency}), and (\ref{eq:cond3}), as follows:
\begin{enumerate}
    \item 
    We adopt the equality in Eq.~(\ref{eq:mu}), which implies that we will obtain a lower limit to the BH and spin;
    \item 
    We replace it into the equality of Eq.~(\ref{eq:Btransparency}), which implies that we are adopting the upper limit to the magnetic field strength (for a given $\alpha$);
    \item
    We obtain the following expression for $\beta$ as a function of $\alpha$ and of the observables $E_{\rm GeV}$ and $L_{\rm GeV}$ \citep{2019ApJ...886...82R}:
\begin{eqnarray}\label{eq:alphabeta}
    \beta&=&\beta(\epsilon_\gamma,E_{\rm GeV},L_{\rm GeV,1},\alpha) \nonumber \\
    &=& \frac{1}{\alpha}\left(\frac{64}{9} \sqrt{3\frac{e^2}{\hbar c}}\frac{\epsilon_\gamma}{B_c^2 r_+(\mu,\alpha)^3}\frac{L_{\rm GeV,1}}{e B_c c^2}\right)^{2/7},
\end{eqnarray}
where we have substituted Eq.~(\ref{maxgev}) into Eq.~(\ref{tcr}) to express $t_c$ as a function of the peak photon energy $\epsilon_\gamma$, instead of the pitch angle $\chi$.
\end{enumerate}

{Therefore,} the BH horizon $r_+$ is a function of $\mu$ and $\alpha$, but in view of Eq.~(\ref{eq:mu}) it becomes a function of $E_{\rm GeV}$ and $\alpha$. Given the observational quantities $E_{\rm GeV}$ (integrated after the UPE phase) and the luminosity $L_{\rm GeV,1}$ (at the end of the UPE phase), Eq.~(\ref{eq:alphabeta}) gives a family of solutions of $\beta$ as a function of $\alpha$. The solution of this equation together with Eq.~(\ref{eq:Btransparency}) gives the values of $\beta$ and $\alpha$. With the knowledge of $\alpha$ and $E_{\rm GeV}$, we obtain $\mu$ from Eq.~(\ref{eq:mu}).

\section{Application to GRB 190114C}\label{sec:8}

\begin{figure}
    \centering
    \includegraphics[width=\hsize,clip]{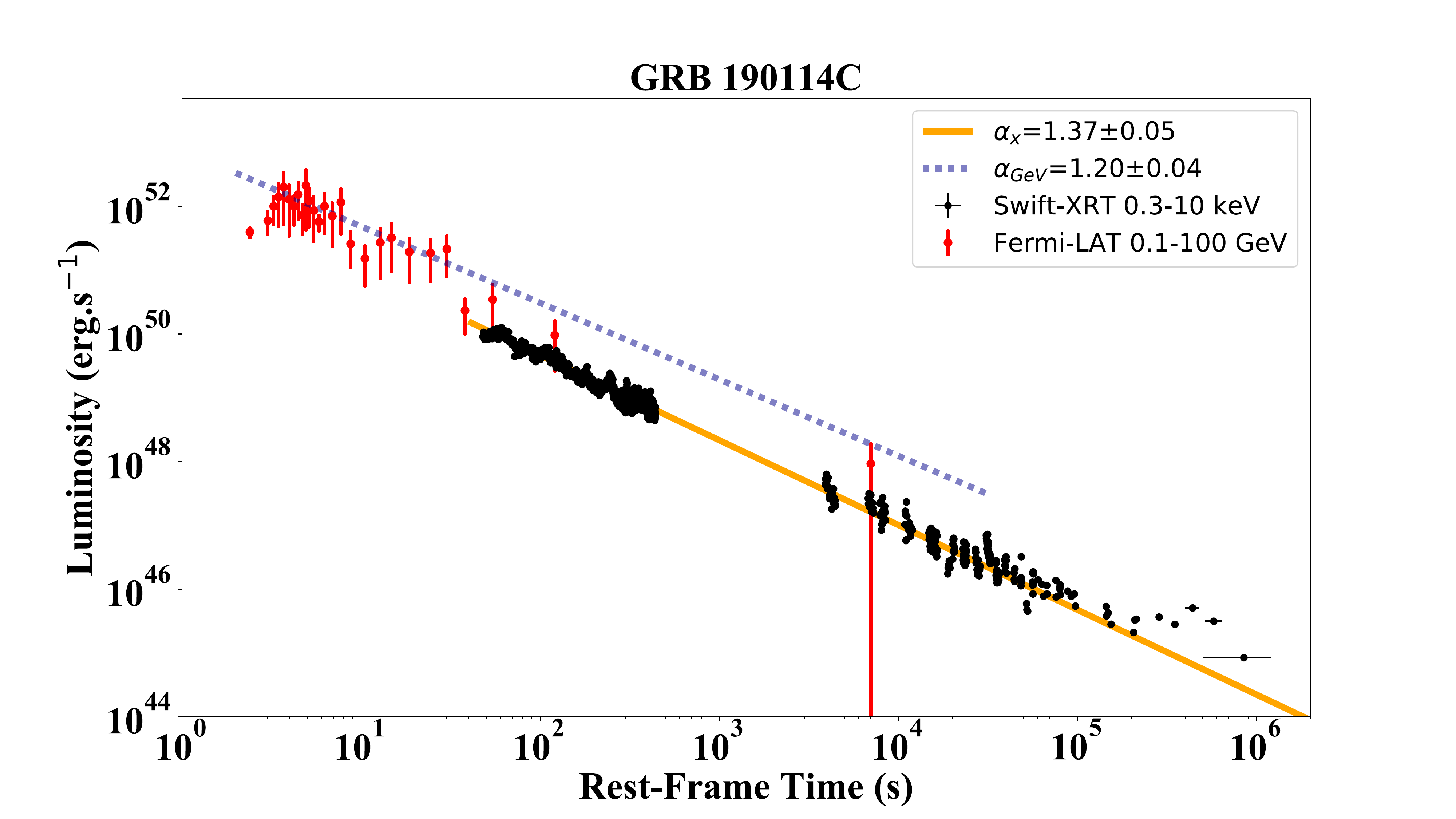}
    \caption{Red data points show the rest-frame $0.1$--$100$~GeV luminosity light curve of GRB 190114C during and after UPE phase obtained from \textit{Fermi}-LAT. The green dashed line shows the best fit for power-law behavior of the luminosity following the UPE phase with slope of $1.2\pm 0.04$ and amplitude of $7.75 \pm 0.44 \times 10^{52}$~erg~s$^{-1}$. {The black data points show the rest-frame $0.3$--$10$~keV luminosity expressed in the rest frame obtained from \textit{Swift}-XRT. It follows a power-law behavior with an amplitude of $A_X = (5.14\pm 2.03)\times 10^{52}$~erg~s$^{-1}$ and a slope of $\alpha_X=1.37\pm 0.05$.}}
    \label{fig:GeVlum}
\end{figure}

\begin{figure}[h]
    \centering
    \includegraphics[width=\hsize,clip]{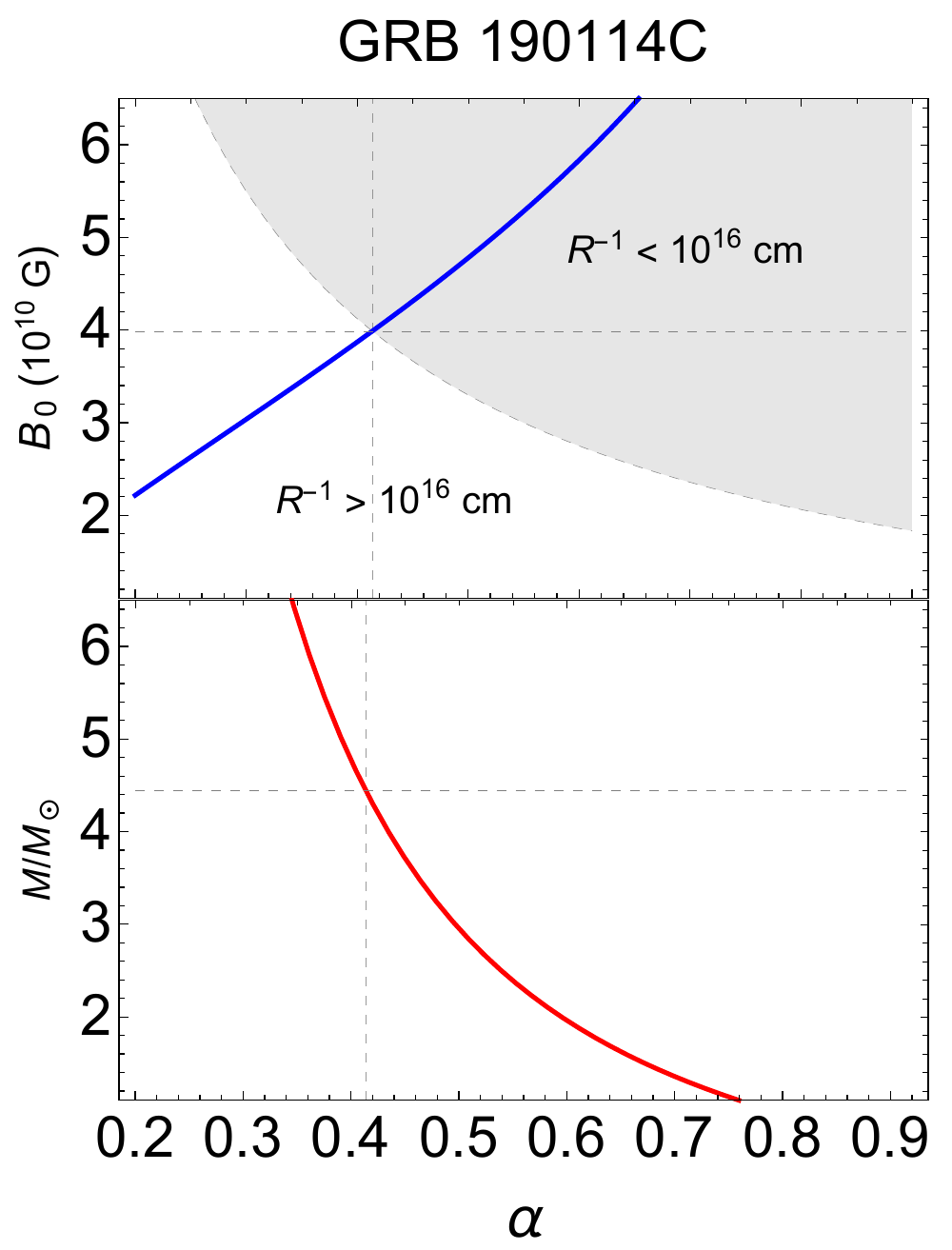}
    \caption{Parameters of the \emph{inner engine} of GRB 190114C. For this source we have $L_{\rm GeV,1}=1.47\times 10^{52}$~erg~s$^{-1}$ and $E_{\rm GeV}=1.8\times 10^{53}$~erg. Upper panel: Family of solutions of $B_0$ as a function of $\alpha$ (blue curve), given by Eq.~(\ref{eq:alphabeta}). We use here a photon energy $\epsilon_\gamma=0.1$~GeV (lower edge of the Fermi-LAT energy band). In the gray shaded region the mean free path is $\bar{R}^{-1} < 10^{16}$~cm, while in the white shaded region it is   $\bar{R}^{-1}\geq 10^{16}$~cm. The curve separating the two regions is therefore given by the equality in (\ref{eq:Btransparency}). Lower panel: Corresponding family of solutions $M(\alpha)$ (red curve), given by Eq.~(\ref{eq:mu}).}
    \label{fig:Bvsalpha}
\end{figure}

We now turn to apply the above procedure to GRB 190114C. For the observational properties of this source we follow the results of \citet{2019arXiv191012615L}. The UPE phase has been  shown in that paper to end at the rest-frame time $t_{\rm rf,UPE}=3.99$~s, so we limit our analysis to longer times. Therefore, for GRB 190114C, we have $t_{\rm rad,1} = t_{\rm rf, UPE} = 3.99$~s. The $0.1$--$100$ GeV luminosity observed by \textit{Fermi}-LAT, at $t \geq t_{\rm rf,UPE}$, is well fitted by a power-law function analogous to the case of GRB 130427A \citep{2019ApJ...886...82R} (see our  Fig.~\ref{fig:GeVlum}):
\begin{equation}\label{eq:Lgev}
    L_{\rm GeV} = A\,t^{-n} = (7.75\pm 0.44) \times 10^{52}\, t^{-(1.2\pm 0.04)}\,\,\text{erg s}^{-1}.
\end{equation}
The associated integrated isotropic energy observed by Fermi-LAT, {from $t_1=t_{\rm rf,UPE}=3.99$~s to $\sim 10^4$~s, is} $E_{\rm GeV}= (1.8 \pm 1.3)\times 10^{53}$~erg, and the luminosity at $t=t_{\rm rad, 1}$ is $L_{\rm GeV,1}=1.47\times 10^{52}$~erg~s$^{-1}$ \citep[see][]{2019arXiv191012615L}.

For the above numbers, and assuming the minimum energy budget requirement, specifically assuming the equality in Eq.~(\ref{eq:Eext}), the \emph{inner engine} parameters are (see Fig.~\ref{fig:Bvsalpha}) magnetic field $B_0\approx 3.9\times 10^{10}$~G, and  spin and BH mass $\alpha = 0.41$ and $M=4.45~M_\odot$, respectively. The corresponding BH irreducible mass is $M_{\rm irr}=4.35~M_\odot$. For the above spin value, Eq.~(\ref{maxgev}) leads to a pitch angle for the emission of $0.1$~GeV photons, $\theta \approx \pi/9$.

The inequality (\ref{eq:Eext}) implies that the above mass and spin values of the BH must be considered as lower limits. As we  show in Sect.~\ref{sec:6}, the \emph{inner engine} can be long-lasting so it can continue to emit and so will eventually radiate more than the observed $E_{\rm GeV}$ that we have used limiting ourselves to the first $10^4$~s of emission. However, in view of the power-law behavior of the GeV luminosity, most of the energy is emitted in this early evolution so the BH parameters do not change significantly if we consider the extrapolation of the energy budget. For example, let us assume that the power-law luminosity (\ref{eq:Lgev}) extends for $1000$~yr. This would increase the total GeV energy radiated by $25\%$, and recalculating all the parameters we obtain $\alpha = 0.46$,  $M=4.50~M_\odot$, and $M_{\rm irr}=4.38~M_\odot$.

\section{{Comparison with previous literature}}\label{sec:9}

\subsection{Long GRBs in the traditional model}\label{sec:9.1}

We first recall some key features of the traditional model of long GRBs. To this end, we are facilitated by the book by Bing Zhang \citep{2018pgrb.book.....Z}, which includes an extensive number of references. The traditional GRB model roots can be found in the papers by \citet{1992MNRAS.258P..41R,1538-4357-482-1-L29}, and \citet{1993ApJ...405..273W}. The model {proposed} by \citet{1992MNRAS.258P..41R} {is based on a single system:} GRBs are explained by a single BH from which an ultrarelativistic blastwave originates and whose expansion follows the Blandford–McKee self-similar solution \citep{1976PhFl...19.1130B}.  \citet{1993ApJ...405..273W} connected the GRB origin to a Kerr BH emitting an ultrarelativistic jet that originates from matter accretion onto the BH. The BH was proposed to form from the direct collapse of a massive star, called  a \emph{failed} SN or  a ``\emph{collapsar},''  leading to a BH in the mass range $5$--$10~M_{\odot}$.

In these models the afterglows are explained via the kinetic energy of the ultrarelativistic blastwave, which can reach very high \emph{bulk} Lorentz factors, $\Gamma \sim 1000$, to be released when it interacts with the circumburst medium \citep{1994ApJ...433L..85W, 1995ApJ...455L.143S, 1997MNRAS.288L..51W, 1997ApJ...489L..37S, 1998ApJ...497L..17S}. The observed spectrum is proposed to be produced by synchrotron and synchrotron self-Compton (SSC) emission from accelerated electrons during the deceleration of the ultrarelativistic blastwave radiating at distances $10^{16}$–$10^{18}$~cm. As pointed out by \citet{2018pgrb.book.....Z}, these models based on a ultrarelativistic blastwave {radiating at distances $10^{16}$–$10^{18}$~cm} have been applied to jointly explain in the jetted emission several observations: 
\begin{enumerate}
    \item the X-ray afterglow, the steep and shallow decay of the luminosity in the \emph{Nousek-Zhang} phase \citep{Nousek2006,Zhang2006}, the X-ray flares and the gamma-ray flares;
 \item {the emission in the MeV and in the keV observed by the Neils Gehrels \textit{Swift} Observatory, as well as the emission in  the optical and in the radio, as well as the emission in the TeV recently observed by MAGIC \citep{GCN23701,2019Natur.575..455M, 2019Natur.575..459M}};
 \item the high-energy (GeV) emission observed in some long GRBs by Fermi-LAT.
\end{enumerate}

{Within the traditional model, all the above emissions are explained using solely the kinetic energy of the ultrarelativistic blastwave with $\Gamma\sim 10^3$, and radiating at $10^{16}$–$10^{18}$~cm. It becomes clear that, requiring to the single kinetic energy of an ultrarelativistic blastwave to account for the entire energetics of all the observed radiation, at all wavelengths, from the prompt to the afterglow, results in an extreme request to the energy reservoir of the GRB engine.}

{
Within the traditional \textit{collapsar}-\textit{fireball} model, the presence of a mildly relativistic
expanding component has been introduced in \citet{2002MNRAS.337.1349R}, called  a \textit{cocoon}, which moves sideways to the jet. However, as  is clearly elucidated in \citet{2017ApJ...834...28N} (see also references therein), the emission in the X-rays from this \textit{cocoon} is too low with respect to be observed X-ray afterglow of long GRBs, unless the \textit{cocoon} Lorentz factor becomes $\Gamma >10$. The possibility of a mildly relativistic component in the traditional model is interesting for its implications for the nature of the low-energy sources such as GRB 060218 \citep[see, e.g.,][]{2015ApJ...807..172N}. These sources have energies $10^{49}$--$10^{51}$~erg, which is the range of energies of BdHNe II and III. However, this is beyond the scope of this article, which is dedicated to BdHNe I, which are characterized by the energy range $10^{52}$--$10^{54}$~erg. In conclusion, an explanation of the X-ray afterglow in the traditional model needs ultrarelativistic values of the Lorentz factor \citep[see also][for a review on the subject]{2018pgrb.book.....Z}
}.

\subsection{Long GRBs in the BdHN model}\label{sec:9.2}

As we  note in Sect.~\ref{sec:1}, the BdHNe have a binary progenitor composed of a CO star and a companion NS. The GRB is composed of independent physical process identified by a time-resolved spectral analysis. Some key results are the following:
\begin{enumerate}
    \item 
   In the analysis of the data of the XRT detector on board the \textit{Neils Gehrels Swift} satellite of the gamma-ray flare, the X-ray flares, the flare-plateau, and the early afterglow phases (the \emph{Nousek-Zhang phase}), after the ultrarelativistic prompt radiation phase, showed that the emitter in these phases is in mildly relativistic expansion with $\Gamma\lesssim 5$ \citep[see][for details]{2018ApJ...852...53R}. A similar upper limit $\Gamma \lesssim 3$ was  obtained in the case of GRB 151027A \citep{2018ApJ...869..151R}, and for GRB 130427A the corresponding upper limit on the \emph{bulk} Lorentz factor is $\Gamma \lesssim 2$ \citep{2018ApJ...869..101R}. Therefore, these stringent upper limits on $\Gamma$ exclude any ultrarelativistic motion following the UPE phase, contrary to the prediction of traditional GRB models based on the ultrarelativistic blastwave.
    \item 
    The  high-energy GeV emission follows from the action of the \textit{inner engine} presented in this work, powered by the BH rotational energy extraction process. In the case of GRB 190114C studied in this work, this corresponds to $t_{\rm rf}\gtrsim 3.99$~s (see Fig.~\ref{fig:GeVlum} and \citealp{2019arXiv191012615L}). It is characterized by an afterglow in the GeV radiation which, when expressed in the rest frame, follows a power-law luminosity (see Eq.~\ref{eq:Lgev} and Fig.~\ref{fig:GeVlum}), and it carries an energy of $E_{\rm GeV}= (1.8 \pm 1.3)\times 10^{53}$~erg.
    \item 
    In parallel, the X-ray afterglow emission observed by the \textit{Swift} satellite originates from the synchrotron radiation produced in the expanding SN ejecta, threaded by the magnetic field of the $\nu$NS, and aided by the injection of particles and the pulsar-like radiation from the $\nu$NS into the SN ejecta \citep{2018ApJ...869..101R, 2019ApJ...874...39W, 2020ApJ...893..148R}. These processes are mainly powered by the rotational energy of the $\nu$NS and have led to a significant progress in  understanding the origin of the X-ray afterglow emission (see, e.g., the case of GRB 130427A in \citealt{2018ApJ...869..101R}, and GRB 160509A, GRB 160625B, GRB 180728A, and GRB 190114C in \citealt{2020ApJ...893..148R}). In these analyses  the spin of the $\nu$NS and the strength and structure of its magnetic field  have been inferred. In the case of GRB 190114C, the luminosity expressed in the rest frame  follows a power-law behavior $L_X = A_X t^{-\alpha_X}$, where $A_X = (5.14\pm 2.03)\times 10^{52}$~erg~s$^{-1}$ and $\alpha_X=1.37\pm 0.05$ and carries an energy $E_X=2.11\times10^{52}$~erg{; see Fig.~\ref{fig:GeVlum}} (\citealp{2021MNRAS.tmp..868R}; see also \citealp{2019arXiv191012615L}). This interpretation of the X-ray afterglow in the BdHN model conforms with the observational upper limits on the $\Gamma$ factor of the X-ray afterglow emitter summarized in point 1 above \citep[see][for details]{2018ApJ...852...53R}.
    \end{enumerate}

In this way, being the total energetics divided into the different components of the system and their associated different physical phenomena, the energetic request to each emission episode in the BdHN becomes affordable.

\subsection{Process of BH energy extraction}\label{sec:9.3}

Having indicated the main differences between the traditional GRB model and the BdHN model regarding the X-ray and the GeV afterglow emissions, we focus now on the mechanism of the high-energy (GeV) emission, which is intimately related to the physics of the GRB central engine.

There is a vast literature devoted to magnetic fields around BHs and how they may act in a mechanism that could extract the rotational energy of a Kerr BH. {An early} attempt in the absence of a charge by a matter-dominated magnetized plasma accreting in a disk around a pre-existing Kerr BH was presented in \cite{1975PhRvD..12.2959R}. The effective potential describing the circular orbit of massive particles around a Kerr BH was adopted (see Ruffini \& Wheeler 1969, in problem 2 of Sect.~104 in \citealp{1975ctf..book.....L}). The infinite conductivity condition, $F_{\mu \nu}u^{\nu}= 0$, where $F_{\mu \nu}$ is the electromagnetic field tensor and $u^{\nu}$ the plasma four-velocity, was  used there leading to $E\cdot B= 0$. Under these conditions, the acceleration of particles and processes of energy extraction were not possible.

This work was further developed by \citet{1977MNRAS.179..433B};  in order to overcome the condition $E\cdot B= 0$ in the magnetosphere, they adopted the concepts of \emph{gaps} and spontaneous $e^+e^-$ pair creation, closely following the seminal ideas of pulsar theory by \citet{1971ApJ...164..529S} and \citet{1975ApJ...196...51R}. They imposed a force-free condition, $F_{\rm \mu \nu}J^{\rm \nu}=0$, where $J^{\rm \nu}$ is the current density, as well as \emph{gaps} outside the BH horizon. The aim was to produce an ultrarelativistic matter-dominated plasma whose bulk kinetic energy could be used to explain the energetics of a jet at large distances from the BH.

There is also another direction in the literature following the work of \citet{1982MNRAS.198..345M}. It extends the work of \citet{1977MNRAS.179..433B} and looks at the problem of matter-dominated accretion in presence of a magnetic field anchored to a rotating surrounding disk. Specifically, they proposed an analogy of a rotating BH immersed in a magnetic field with a rotating conductive sphere and/or with the analogy of such a BH and the surrounding magnetosphere as an \emph{electric circuit}. Independent of the analogies, the underlying physical system remains the same as that proposed by  \citet{1977MNRAS.179..433B}.

The present model is {mainly} motivated by fitting the GeV emission of GRBs. There is no matter-dominated disk accretion. There is instead a very low-density ionized plasma fulfilling an acceleration electrodynamical process around a newly born BH. We use the Papapetrou-Wald solution in which the electromagnetic field is naturally characterized by regions where $E\cdot B\neq 0$ (see Sect.~\ref{sec:2}, Fig.~\ref{fig:fieldlines}, and \citealp{1974PhRvD..10.1680W}). This feature naturally allows the acceleration of particles without the need of introducing any \textit{gaps}. There is no ultrarelativistic matter-dominated plasma outflow. The accelerated charged particles emit synchrotron-radiation photons that carry off energy and angular momentum close to the BH. The BH in our scenario is not pre-existing: it is smoothly formed by the hypercritical accretion onto the binary companion NS. The magnetic field, characterizing the Papapetrou-Wald solution, is amplified during the process of gravitational collapse of the binary companion NS \citep{2020ApJ...893..148R}. There is no room in this model for   the gravitational stable circular orbits around the Kerr BH. The particles are accelerated by an ultrarelativistic electrodynamical process.

Our description is also different with respect to recent GRB literature. For instance, in \citet{2011MNRAS.413.2031M}, \citet{2017MNRAS.472.3058B}, and references therein, the presence of a magnetized wind, outflow, or jet is powered by a central engine. In these works the engine is represented by a NS endowed with an ultrahigh magnetic field, a \emph{magnetar}, that loses its rotational energy via magnetic-dipole braking, in complete analogy to pulsars. The \emph{magnetar} powers the outflows that produce the GRB emission at large radii on the order of $10^{15}$~cm. These models focus on the explanation of the (MeV) GRB prompt and the (X-ray) afterglow emission using the rotational energy of a \emph{magnetar}, so they do not look either to the physics of BHs, or to the GeV emission that are the topics of the present article.

The understanding of the {complex} nature of a BdHN requires the knowledge of different episodes, which in some cases are strictly independent, and their description can occur independently of each other.

For instance, the existence of hyper-energetic SN, the \emph{SN-rise}, radiates off $10^{52}$~erg in the case of GRB 190114C \citep{2019arXiv191012615L}. In parallel, the interaction of the SN ejecta with the magnetic field of the $\nu$NS and its pulsar-like emission, explain the observed X-ray afterglow \citep{2018ApJ...869..101R, 2019ApJ...874...39W, 2020ApJ...893..148R}. This emission is produced at distances $10^{12}$--$10^{16}$~cm from the binary progenitor.

In the present work we address the most energetic GRB component, the GeV emission originating close to the horizon, at distances of $10^{6}$~cm, starting in the case of GRB 190114C at a rest-frame time of $3.99$~s after the trigger.

After the clarification of these concepts, we will be ready to describe the optically thick sub-MeV emission in the time interval $1.99$--$3.99$~s, which comprises the $55\%$ of the energy of GRB 190114C, overcoming the compactness problem using our classic approach of the fireshell model (see \citealp{1999A&AS..138..511R,2000A&A...359..855R, 2001A&A...368..377B} and R.~Moradi, et al., in preparation).

\section{Conclusions}\label{sec:10}

The \textit{inner engine} theory   applied in this work to GRB 190114C represents an authentic full paradigm shift  from the traditional model of long GRBs based on the emission of an ultrarelativistic blastwave, somehow powered by a Kerr BH. It seems too expensive for nature to accelerate matter in bulk, against the gravitational pull of the BH, to a large distance of $\sim 10^{16}$--$10^{17}$~cm and with $\Gamma \sim 10^3$ to guarantee the transparency of high-energy radiation. For instance, the explanation of the GRB 190114C high-energy emission needs an ultrarelativistic blastwave with a kinetic energy on the order of $10^{55}$~erg \citep[see, e.g.,][]{2019Natur.575..455M, 2019Natur.575..459M}. It is clear that such  energy cannot be powered by extracting the rotational energy of a Kerr BH of a few $M_\odot$, which will be a few $10^{53}$~erg (see Eq.~\ref{eq:mu}).

We have shown that the \textit{inner engine} can nicely explain the GeV emission by accelerating electrons in the vicinity of the Kerr BH, which radiate their kinetic energy gain via synchrotron emission. The number of particles needed by the \textit{inner engine} to explain the observed high-energy emission is relatively low. Let us adopt the derived \textit{inner engine} parameter for GRB 190114C: $M = 4.4\,M_\odot$, $\alpha = 0.4$, and $B_0 = 4\times 10^{10}$~G. For instance, from Eq.~(\ref{maxgev}) we obtain that for this $\alpha$ a photon peak energy of $10$~GeV is obtained for an electron pitch angle $\chi \approx 0.2^\circ$ (see also Fig.~\ref{fig:chivsegamma}). Using Eq.~(\ref{gas}), this implies an electron Lorentz factor $\gamma \approx 6.76\times 10^4$, which corresponds to an electron energy $\epsilon_e = \gamma m_e\,c^2 \approx 5.53\times 10^{-2}$~erg $= 3.45\times 10^{10}$~eV. Therefore, the number of such electrons needed to power the GeV emission of total energy $E_{\rm GeV} = 1.8\times 10^{53}~{\rm erg}\approx 0.1~M_\odot c^2$, is $N_e = E_{\rm GeV}/\epsilon_e = 3.25\times 10^{54}$, which for ionized matter implies a mass of $m_p N_e \approx 2.73\times 10^{-3}~M_\odot$, where $m_p$ is the proton mass.

Therefore, the \emph{inner engine} uses a more efficient electrodynamical process that produces observable high-energy emission in the vicinity of the BH. In fact the acceleration is not based on a bulk-expanding motion. Every single electron is accelerated from its initial velocity up to an asymptotic value defined by the maximum electric potential energy available for their acceleration, which depends only on the external magnetic field strength and the BH spin parameter; see Eq.~(\ref{eq:deltaphi}). These accelerated electrons radiate mainly at high energies in the GeV domain. The radiation of the \textit{inner engine} (e.g., at keV to MeV energies) is negligible (with respect to the observed values). The observed radiation in the keV to MeV energy domains is explained by a different mechanism in a BdHN I; see \citep{2020ApJ...893..148R}. The observed luminosity of GeV allows us to estimate the mass and spin of the BH.

We have determined the parameters of the \emph{inner engine} of GRB 190114C using only the GeV emission data after the UPE phase. We asked the system to satisfy three physical conditions. First, that the GeV energetics is paid by the extractable energy of the BH (see Eq.~(\ref{eq:mu})); second that the system is transparent to GeV photons produced by the synchrotron radiation of the accelerated electrons (see Eq.~(\ref{eq:Btransparency})); and third that     the synchrotron radiation timescale explains the observed GeV emission timescale (see Eq.~(\ref{eq:cond3})) with the aid of Eq.~(\ref{tcr}). In order to be fulfilled, this last constraint implies that the GeV emission is emitted from electrons being accelerated with the appropriate pitch angles ({see Figs.~\ref{fig:chivsegamma} and \ref{fig:pitchangles}}). These pitch angles occur within a cone of approximately $60^\circ$ from the BH rotation axis ({see Fig.~\ref{fig:pitchangles}}), which is a key result for the interpretation of the morphology of the BdHN I \citep{2021MNRAS.tmp..868R}.

From this procedure, we have obtained the inner engine parameters of GRB 190114C: $B_0\approx 3.9\times 10^{10}$~G, $\alpha\approx 0.41$, and $M=4.45~M_\odot$. The corresponding irreducible mass of the BH is $M_{\rm irr}=4.35~M_\odot$. {It is worth recalling that both $M_{\rm irr}$ and $B_0$ are kept constant and this should be all over the evolution.} {The corresponding BH parameters for GRB 130427A are dimensionless spin $\alpha= 0.47$, mass $M=2.3~M_\odot$, and irreducible mass $M_{\rm irr}=2.2~M_\odot$ \citep{2019ApJ...886...82R}.} The above are the first two BH masses derived directly from the GRB observations, and in both cases they are above the theoretical values of the NS critical mass enforcing the validity of the BdHN I model: the BH are formed by smooth hypercritical accretion of the HN ejecta on the NS binary companion.

Since here we only  used  the GeV emission data, the BH parameters that we have obtained, namely mass and spin, have to be considered as lower limits. Thus, it is clear that even a slightly higher mass (or spin) of the BH can guarantee even larger and longer emission of the \textit{inner engine}.
 
Our analysis paves the way to additional research; the data from the different energy bands ({e.g.,} the higher energy bands; \citealt{2019Natur.575..455M, 2019Natur.575..459M}) {might provide additional} information on the energy distribution of the electrons injected by the electric field into the magnetic field, and on the pitch angle distribution for the synchrotron emission. Figure~\ref{fig:pitchangles} shows, for the electromagnetic field configuration of the Papapetrou-Wald solution (see Fig.~\ref{fig:fieldlines}), the contours of constant pitch angle and constant electric energy density.

Before concluding, it is worth   recalling some crucial aspects of the \emph{inner engine} here applied to the case of GRB 190114C. The nature of the emission results from considering the physical process leading to the electric and magnetic fields and the BH formation (see Sect.~\ref{sec:3} and \citealp{2020ApJ...893..148R}). This is fundamental to show that the emission process leading to the observed luminosity is not continuous, but discrete. The timescale of the emission in GRBs is too short to be probed directly by current observational facilities. Direct evidence of the value and discreteness {might} come   instead from the observation of large BHs of $10^8$--$10^{10}~M_\odot$ in AGN. For instance, {in the case of M87*, for fiducial parameters $M = 6\times 10^9~M_\odot$, $\alpha = 0.1$, and $B_0=10~G$, the \textit{inner engine} theory predicts a high-energy (GeV) emission with a luminosity of a few $10^{43}$~erg~s$^{-1}$, with a timescale of up to tenths of seconds (see Table~\ref{tab:parameters}). Emission at higher energies (e.g., in the TeV band), would be characterized by a lower luminosity and a longer timescale. The timescale for UHECR emission is instead approximately half a day (see Table~\ref{tab:parameters} and \citealp{2020EPJC...80..300R})}.

We can therefore conclude, in the light of the results of this article and the previous articles in this series, that all BdHN I are powered by three independent sources of energy. The BdHN I is triggered by the SN explosion originating from the collapse of the CO$_{\rm core}$ generating a $\nu$NS. The accretion of the SN onto the $\nu$NS (see Sect.~\ref{sec:9.2} and \citealp{2018ApJ...869..101R, 2019ApJ...874...39W, 2020ApJ...893..148R}), gives origin to the X-ray afterglow observed by \emph{Swift}. The hypercritical accretion of the SN onto the binary companion NS gives origin to the BH as soon as the NS reaches the critical mass. This smooth accretion process is alternative to the direct gravitational collapse of a massive star. This happens in GRB 190114C at $t_{\rm rf} = 1.99$~s. The further accretion of the SN ejecta onto the newly born BH generates the prompt gamma-ray radiation observed in GRB 190114C between $1.99$~s and $3.99$~s (R.~Moradi, et al., in preparation). The further accretion of the SN ejecta onto the newly born BH leads to a process of energy extraction from the \emph{inner engine} that generates the jetted high-energy ($\gtrsim$GeV) emission. This radiation, as is shown in this article using the Papapetrou-Wald solution (see Sect.~\ref{sec:2}), is emitted close to the BH horizon and within an angle of nearly $60^\circ$ from the BH rotation axis (see Sect.~\ref{sec:5} and Fig.~\ref{fig:pitchangles}).

\begin{acknowledgements}
We thank the Editor and the Referee for the reiterated suggestions which have certainly improved {the presentation of our results. An extended correspondence with the Referee has been addressed since our initial submission in order to broader the context of our work with respect to the previous literature and which have materialized in Sect.~\ref{sec:9}.}
\end{acknowledgements}


\end{document}